\documentclass[structabstract]{aa}  
\usepackage{graphicx}
\usepackage{natbib}
\bibpunct{(}{)}{;}{a}{}{,}
\begin{document}
   \title{Observation of SN2011fe with \emph{INTEGRAL}. I. Pre--maximum phase\thanks{ Based on observations with \emph{INTEGRAL}, an ESA project with instruments and science data centre funded by ESA member states (specially the PI countries: Denmark, France, Germany, Italy, Switzerland, and Spain), the Czech Republic, and Poland and with the participation of Russia and USA.}
} 



    \author{J. Isern\inst{1} 
\and P. Jean \inst{2,3} 
\and E. Bravo \inst{4} 
\and R. Diehl \inst{5}
\and J. Kn\"odlseder\inst{2,3} 
\and A. Domingo \inst{6} 
\and A. Hirschmann \inst{1}
\and P. Hoeflich \inst{7} 
\and F. Lebrun \inst{8}
\and M. Renaud \inst{9}
\and S. Soldi \inst{10}
\and N. Elias--Rosa \inst{1} 
\and M. Hernanz \inst{1}
\and B. Kulebi \inst{1}
\and X. Zhang \inst{5}
\and C. Badenes \inst{11}
\and I. Dom\'{\i}nguez  \inst{12}
\and D. Garcia-Senz \inst{4}
\and C. Jordi \inst{13} 
\and G. Lichti \inst{5}
\and G. Vedrenne \inst{2,3}
\and P. Von Ballmoos \inst{2,3}
}

   \institute{Institut de Ci\`encies de l'Espai (ICE-CSIC/IEEC),
             Campus UAB, 08193 Bellaterra, Barcelona, Spain\\
              \email{nelias@ieec.cat};\email{hernanz@ieec.cat};\email {hirschmann@ieec.cat}; 
              \email{isern@ieec.cat};  \email{kulebi@ieec.cat}
\and Universit\'e de Toulouse; UPS-OMP; IRAP; Toulouse, France\\
\and IRAP, 9 Av colonel Roche, BP44346, 31028 Toulouse Cedex 4, France\\
       \email{pierre.jean@irap.omp.eu}; \email{jknodlsede@irap.omp.eu}; \email{pvb@cesr.fr};
       \email{vedrenne@cesr.fr}
\and Dept. Fisica i Enginyeria Nuclear, Univ. Polit\`{e}cnica de Catalunya,
                          Barcelona, Spain\\
                           \email{eduardo.bravo@upc.edu}; \email{domingo.garcia@upc.edu}
\and Max -Planck-Institut for Extraterrestrial Physics, 
               Giessenbachstrasse 1, D-85741 Garching , Germany\\
       \email{rod@mpe.mpg.de}; \email{grl@mpe.mpg.de}; \email{zhangx@mpe.mpg.de}
\and Centro de Astrobiolog\'{\i}a (CAB-CSIC/INTA),
           P.O. Box 78, 28691 Villanueva de la Ca\~{n}ada, Madrid, Spain \\
       \email{albert@cab.inta-csic.es}
\and Physics Department, Florida State University,
        Tallaharssee, FL32306, USA \\
       \email{pah@aastro.physics.fsu.edu}
\and Astroparticle et Cosmologie (APC), CNRS-UMR 7164, Universit\'e de Paris 7 Denis Dioderot,
        F-75205 Paris, France
       \email{lebrun@apc.univ-paris7.fr}
\and Laboratoire Univers et Particules de Montpellier (LUPM), UMR 5299, Universit\'e de Montpellier II, F-34095, Montpellier, France \\
        \email{mrenaud@lupm.univ-montp2.fr}
\and AIM (UMR 7158 CEA/DSM-CNRS-Universit\'e Paris Diderot),
       Irfu/Service d'Astrophysique, F-91191 Gif-sur-Yvette, France \\
        \email{simona.soldi@cea.fr}
\and Department of Physics and Astron omy \& Pittsburgh Particle Physics, Astrophysics and
        Cosmology Center (PITT-PACC), University of Pittsburgh, Pittsburgh PA15260, USA \\
        \email{badenes@pitt.edu}
\and Universidad de Granada, C/Bajo de Huetor 24, Apdo 3004, E-180719, Granada, Spain \\
        \email {inma@ugr.es}
\and Dept. d'Astronomia i Meteorologia, Institut de Ci\`encies del Cosmos (ICC), Universitat de Barcelona (IEEC-UB) \\
        \email {carme.jordi@ub.edu}
            }

   \date{\today}

\abstract{
SN2011fe was detected by the Palomar Transient Factory on August 24th 2011 in M101 a few hours after the explosion. From the early optical spectra it was immediately realized that it was a Type Ia supernova thus making this event the brightest one discovered in the last twenty years.
}
{ The distance of the event offered the rare opportunity to perform a detailed observation with the instruments on board of \emph{INTEGRAL} to detect the $\gamma$--ray emission expected from the decay chains of $^{56}$Ni. The observations were performed in two runs, one before and around the optical maximum, aimed to detect the early emission from the decay of $^{56}$Ni and another after this maximum aimed to detect the emission of $^{56}$Co.
} 
{ The observations performed with the instruments on board of \emph{INTEGRAL} (SPI, IBIS/ISGRI,  JEMX and OMC) have been analyzed and compared with the existing models of $\gamma$--ray emission from such kind of supernovae. In this paper, the analysis of the $\gamma$--ray emission has been restricted to the first epoch.}
{ Both, SPI and IBIS/ISGRI, only provide upper-limits to the expected emission due to the decay of $^{56}$Ni. These upper-limits on the gamma-ray flux are of 7.1 $\times$ 10$^{-5}$ ph/s/cm$^2$ for the 158 keV line and of 2.3 $\times$ 10$^{-4}$ ph/s/cm$^2$ for the 812 keV line. These bounds allow to reject at the $2\sigma$ level
 explosions involving a massive white dwarf, $\sim 1$ M$\odot$  in the sub--Chandrasekhar scenario and specifically all models that would have substantial amounts of radioactive $^{56}$Ni in the outer layers of the exploding star responsible of the SN2011fe event. The optical light curve obtained with the OMC camera also suggests that SN2011fe was the outcome of the explosion, possibly a delayed detonation although other models are possible, of a CO white dwarf that synthesized $\sim 0.55$ M$_\odot$ of $^{56}$Ni. For this specific model, INTEGRAL would have only been able to detect this early $\gamma$--ray emission if the supernova had occurred at a distance $\la 2$ Mpc.}
{The detection of the early $\gamma$--ray emission of $^{56}$Ni is difficult and it can only be achieved with \emph{INTEGRAL} if the distance of the event is close enough. The exact distance depends on the specific SNIa subtype. The broadness and rapid rise of the lines are probably at the origin of such difficulty.}

  
   \maketitle
%

\section{Introduction}

From the photometric point of view, Type Ia supernovae (SNIa) are characterized 
by a sudden rise and decay of their luminosity, followed by a slowly-- fading tail. From the spectroscopic point of view, they are characterized by the lack of H--lines and the presence of Si II--lines in their spectra during the maximum light and by the presence of Fe emission features during the nebular phase. A noticeable property is the spectrophotometric homogeneity of the different outbursts. Furthermore, in contrast with the other supernova types, they appear in all kind of galaxies. These properties point out an exploding object that is compact, free of hydrogen, that can be activated on short and long time scales, and is able to synthesize a minimum of 0.3 M$_\odot$ of radioactive \element[ ][56]{Ni} to power the light curve. These constraints immediately led to the proposal that SNIa were the outcome of the thermonuclear explosion of a mass accreting C/O white dwarf (WD) near the Chandrasekhar's limit \citep{hoyl60} in a close binary system. 

Despite this homogeneity, when SNIa are observed in detail some  differences appear. Now it is
known that there is  a group of  SNIa with light  curves showing very  bright and
broad peaks,  the SN1991T  class, that represents  9\% of  all the
events. There  is another  group with a much dimmer and narrower peak and that
lacks of the  characteristic secondary peak in the infrared, the SN1991bg  class, that
represents 15\% of all the  events. To these categories it has been
recently added a  new one that contains very  peculiar supernovae, the
SN2002cx class,  representing $\sim 5$\%  of the  total. These
supernovae are characterized by high ionization spectral features in
the pre-maximum, like the SN1991T  class, a very low luminosity,
and  the lack  of a secondary  maximum in the infrared,  like the  SN1991bg class.   The
remaining ones, which amount  to $\sim 70\%$, present normal behaviors and are  known as 
\emph{Branch-normal} \citep{li11a}. However, even the  normal ones
are  not completely  homogeneous  and show  different luminosities  at
maximum and light curves with different decline rates \citep{li11b}. This variety has recently increased with the discovery of SN2001ay, which is characterized by a fast rise and a very slow decline \citep{baro12}.
This diversity strongly suggests that different scenarios and burning 
mechanisms could be operating in the explosion.

From the point of view of the explosion mechanism as seen in one dimensional models, it is
possible to distinguish four cases \citep{hoef96,hill00}: the pure detonation model (DET), the pure
deflagration model (DEF), the delayed detonation model (DDT), and the pulsating detonation 
model (PDD). The equivalent models in three dimensions also exist, but with a larger variety of 
possibilities. An additional class  are the so called Sub--Chandrasekhar's (SCh) models 
in which a detonation triggered by the ignition of He near the base  of a freshly accreted 
helium layer completely burns the white dwarf. At present, there is no basic argument to reject any of the models, except the DET ones that are incompatible with the properties of 
the spectrum of SNIa at maximum light. Present observations also pose severe constraints to 
the total amount of \element[ ][56]{Ni} that can be produced by the He--layer in SCh models 
\citep{hoef96,nuge97,woos11}.

According to the nature of the companion, either non--degenerate or degenerate, progenitors can be classified as single degenerate systems --SD \citep{whel73} or double degenerate systems --DD \citep{webb84,iben85}. The distinction among them is important in order to interpret the observations since, depending on the case, the white dwarf can ignite below, near or above the Chandrasekhar's mass and consequently the total mass ejected and the mass of \element[ ][56]{Ni} synthesized can be different. At present, it is not known if both scenarios can coexist or just one is enough to account for the supernova variety. Observations of the stellar content in the interior of known SNIa remnants point towards one possible SD candidate \citep[Tycho SNR, see][]{ruiz04,her09,ker09} and two almost certain DD candidates \citep[SNR0509-67.5 and SNR 0519-69.0][]{sch12,edw12}.

The detection of $\gamma$--rays from supernovae can provide important insight on the nature of the progenitor and especially on the explosion mechanism, since the amount and distribution of the radioactive material produced in the explosion strongly depend on how the ignition starts and how the nuclear flame propagates \citep[see][for a detailed discussion of how these differences are reflected in the spectra]{gome98,iser08}. The advantages of using $\gamma$--rays for diagnostic purposes relies on their penetrative capabilities, on their ability to distinguish among different isotopes and on the relative simplicity of their transport modelling as compared with other regions of the electromagnetic spectrum. Unfortunately, such observations have not bee achieved so far because of the poor sensitivity of the instruments. For this reason up to now it has only been possible to place upper limits to the SN1991T \citep{lich94} and SN1998bu \citep{geor02} events. 

Several authors have examined the $\gamma$--ray emission from SNIa \citep{gehr87,ambw88,burr90,ruiz93,hoef94,kuma97,timm97,gome98,sim08}. To explore the above model variants we have used as a guide the properties obtained with the code described in \cite{gome98}, which is based on the methods described by \cite{pozd83} and \cite{ambw88}. In order to test the consistency of this model, the results of this code were successfully cross--checked with those obtained by other authors \citep{miln04}. This code was later generalized to three dimensions \citep{hirs09}.

 Before and around the epoch of maximum of the optical light curve, the $\gamma$--ray emission can be characterized (figure 2 of G\'omez-Gomar et al. loc.cit.)  as follows: i) A spectrum dominated by the \element[ ][56]{Ni} 158 and 812 keV lines. ii) Because of the rapid expansion, the lines are blueshifted but their energy peak quickly evolves back to the red as matter becomes more and more transparent. The emergent lines are broad, typically from  3\%  to 5\%. Because of the Doppler effect the 812 keV line blends with the quickly growing \element[ ][56]{Co} 847 keV line, forming a broad feature. iii) The intensity of the \element[ ][56]{Ni} lines rises very quickly with time, being very weak at the beginning, even in the case of Sub-Chandrasekhar models. This fact, together with the relatively short lifetime of \element[ ][56]{Ni}, makes the observational window rather short.

SN 2011fe (RA = 14:03:05.81, Dec = +54:16:25.4; J2000) was discovered in M101 on August 24th, 2011 \citep{nuge11}. The absence of hydrogen and helium, coupled with the presence of silicon in the spectrum clearly indicates that it belongs to the SNIa class. Since it was not visible on August 23th, this supernova must have been detected $\sim 1$ day after the explosion \citep{nuge11}. Furthermore, as M101 is at a distance of 6.4 Mpc \citep{stet98,shap11}, SN2011fe is the brightest SNIa detected in the last 25 years. This distance is slightly less than the maximum distance at which current gamma-ray instruments should be able to detect an intrinsically luminous SNIa. The closeness of SN2011fe has made it possible to obtain the tightest constraints on the supernova and its progenitor system to date in a variety of observational windows. Red giant and helium stars companions, symbiotic systems, systems at the origin of optically thick winds or containing recurrent novae are excluded for SN2011fe \citep{li11c,bloo12,bro11,cho12}, leaving only either DD or a few cases of SD as possible progenitor systems of this supernova. 

In this study we analyze the data obtained by \emph{INTEGRAL}  during the optical pre--maximum observations spanning from 4258.8733 IJD (August 29th, 20:59 UT) to 4272.1197 IJD (September 12th, 2:52 UT) with a total observation time of 975419 s. This schedule was essentially determined by the constraints imposed by the Sun, according to the TVP tool of \emph{INTEGRAL}, that  prevented the observation just beyond the optical maximum where the \element[ ][56]{Ni} lines are expected to peak. Despite this limitation, these early observations were triggered to constrain any predicted early gamma-ray emission as may be expected from some variants of sub-Chandrasekhar models.  In the next section, we present the data obtained with the instruments onboard \emph{INTEGRAL}. Then, we discuss the limits they can put on present models of SNIa explosion, and conclude.

\section{\emph{INTEGRAL} data}

\emph{INTEGRAL}  \citep{wink03} is able to observe in gamma--rays, X--rays and visible light. 
It was launched in October 17th 2002 and was injected into a highly eccentric orbit with a 
period of about 3 days in such a way that it spents most of its time well outside the radiation 
belts of the Earth. The spacecraft contains two main instruments, SPI, a germanium spectrometer 
for the energy range of 18 keV to 8 MeV with a spectral resolution of 2.2 keV at 1.33 MeV 
\citep{vedr03}, and IBIS, an imager able to provide an imaging resolution of 12 arcmin FWHM \citep{uber03}, which 
has a CdTe detector, ISGRI, able to provide spectral information of the continuum and broad 
lines in the range of 15 keV to 1 MeV \citep{lebr03}. Other instruments onboard are an X--ray monitor, JEM-X, that works in the range of 3 to 35 keV \citep{lund03}, and an optical camera, OMC, able to operate in the visible band of the spectrum up to a magnitude of 18 \citep{mash03}.
  
\subsection{OMC data}

   \begin{figure}
   \centering
 \includegraphics[width=11cm]{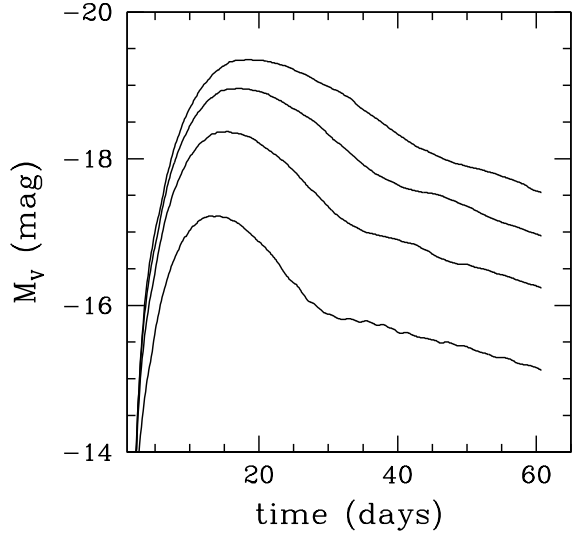}
      \caption{Light curves in the V-band obtained from delayed detonation models that produce, up to down, 0.63, 0.45, 0.27, 0.09 M$_\odot$ of \element[ ][56]{Ni} and satisfy the brightness-decline relationship.
              }
         \label{omv1}
   \end{figure}

The height of the maximum of the optical light curve depends on the total amount of  
\element[ ][56]{Ni} synthesized during the explosion, its distribution in the debris, the total kinetic energy of matter and the opacity \citep{arne96}. The brightness decline relation ( $\Delta m_{\rm 15}$) relates the absolute brightness at maximum light and the  rate of the post-maximum decline over 15 days. From theory, this relationship is well understood: light curves are powered by the radioactive decay of \element[ ][56]{Ni} \citep{colg69}.  More \element[ ][56]{Ni} increases the luminosity and causes the envelopes to be hotter. Higher temperature means higher opacity and, thus, longer diffusion time scales and slower decline rates after maximum light \citep{hoef96a,nuge97,umed99,kase09}. The $\Delta m_{\rm 15}$-relation holds up for virtually all explosion scenarios as long as there is an excess amount of stored energy to be released \citep{hoef96a,baro12}. The tightness of the relation observed for Branch-normal SNe~Ia is about 
 $0.3^m$ \citep{hamu96,perl99} and it is consistent with explosions of models of similar mass. Since delayed detonation models (see Fig. \ref{omv1}) provide a reasonable spectral evolution, their maximum brightness is consistent with the Hubble constant and the $\Delta m_{\rm 15}$-relationship is consistent with the observations with a factor between 1 and 1.3, we will use these models to estimate the total amount of \element[ ][56]{Ni} freshly synthesized.

   \begin{figure}
   \centering
 \includegraphics[width=11cm]{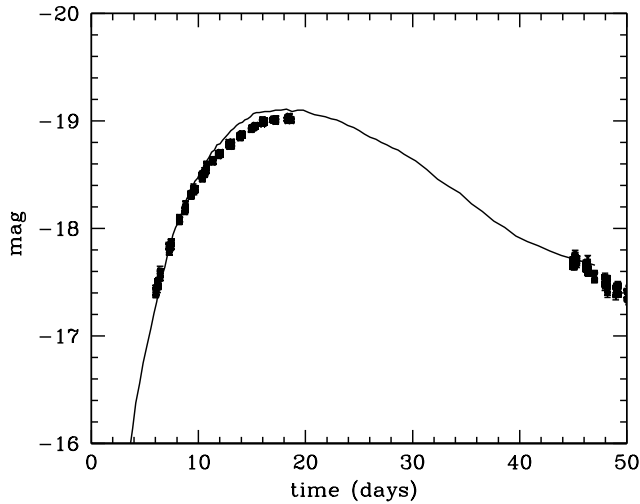}
      \caption{Light curve of SN2011fe in the V--band obtained with the OMC/\emph{INTEGRAL} 
assuming a distance of 6.4 Mpc (dots) for the orbits 1086-1088 and 1097-1101. The continuous line represents a reasonable fit obtained with a DDT model (see text).
              }
         \label{omc2}
   \end{figure}

The properties and the photometric characterization of the OMC can be found in  
\cite{mash03}.  The data obtained during the orbits 1084--1088 plus the data 
corresponding to orbits 1097--1111 were analysed with the Offline Scientific Analysis Software (OSA, version 9) provided by the ISDC Data centre for Astrophysics \citep{cour03}. 
The fluxes and magnitudes were derived from a photometric aperture of $3 \times 3$ pixels (1 pixel = 17.504 arcsec), slightly circularized, i.e. removing 1/4 pixel
from each corner (standard output from OSA). The default centroiding algorithm
was used, i.e. the photometric aperture was centred at the source coordinates.
We checked that the photometric aperture of $3 \times 3$ pixels does not include any significant
contribution by other objects. Because the source was bright enough, combination 
of several shots were not required to increase the signal-to-noise ratio. To only 
include high-quality data, some selection criteria were applied to individual 
photometric points removing those measurements with a problems flag. Shots were 
checked against saturation, rejecting those with long exposures (200 seconds) for $V < 10.5$. 
To avoid noisy measurements the shortest exposures (10 seconds) were not used.

The extinction along the line of sight has two components, one due to the Milky Way and other to M101. Since M101 has a galactic latitude of $\sim 60\degr$ the Milky Way contribution is expected to be small. \cite{shap11} estimate $E(B-V) =0.009$ which means an extinction of $A_\mathrm{V} \sim 0.03$ mag. The M101 contribution is also expected to be small since it is a face on galaxy and SN2011fe is placed in a region with a small concentration of interstellar dust and gas \citep{suzu07,suzu09}. It is possible to obtain a rough upper limit of the extinction using the observations \citep{shap11} of two regions containing Cepheids that overlap just at the position of SN2011fe. The average of the two extinctions, $E(B-V) = 0.2$ mag, would imply an absorption of $A_\mathrm{V} = 0.62$ mag. However, since the region containing the supernova has a small concentration of interstellar matter, the comparison of the light curve at different bands and the absence of strong sodium lines in the spectrum suggest that extinction is small. Thus the adopted absorption affecting the supernova is taken to be $ A_\mathrm{V} \sim 0.03$. 

The V--band light curve reached the maximum, 
$ V = 9.99$, at the day $IJD = 4271.44$, in agreement with \cite{rich12}. 
Taking a canonical distance of 6.4 Mpc (see however \cite{tamm11}) and assuming no extinction,
the absolute magnitude should be $ M_{\rm V} = -19.04 $ at maximum thus 
indicating that the SN2011fe was a slightly dim average SNIa. The V--band 
light curve can be well fitted with a delayed detonation model 
(see Fig.~\ref{omc2}) of a Chandrasekhar mass WD igniting at 
$\rho_\mathrm{C} = 2 \times 10^9\mathrm{ g\, cm}^{-3}$, making 
the transition deflagration/detonation at 
$\rho_\mathrm{tr} = 2.2 \times 10^7 \mathrm{ g\, cm}^{-3}$.  
This model produces 0.51 M$_\odot$ of \element[][56]{Ni}, although 
if extinction and distance uncertainties are taken into account 
this value could easily be a $\sim10\%$ larger \citep{hoef02}. This value is also consistent with the  estimations found by \cite{roep12} using an independent DDT model and a violent merger model and is roughly equivalent to the one obtained with model DDTe of 
Table \ref{table:2}. We note that this yield of \element[][56]{Ni} 
is consistent with the value derived by \citet{nuge11a}. From this 
theoretical model, $\Delta m_\mathrm{15} (B) = 1.2 \pm 0.2$, in agreement 
with \cite{tamm11} but slightly smaller than the value found by \cite{rich12}. 
The observation of the early light curve, 4 hours \citep{bloo12}, 
and 11 hours \citep{nuge11a} after the explosion strongly supports 
the scenario based on the explosion of a C/O white dwarf.

\subsection{SPI data}
 Only data of the low energy range ($\la $ 2 MeV) have been analysed in this study of the SPI output. The energy calibration 
was performed for each orbit by fitting parameters of a four degree polynomial function with the 
channel positions of the instrumental background lines at 23.4 keV, 198.4 keV, 309.9 keV, 584.5 keV, 
882.5 keV and 1764.4 keV. The precision of the resulting calibration is better than $\sim$ 0.1 keV 
at 1 MeV. The calibrated single-detector and multiple-detector\footnote{We used only double-detector 
events as including events with larger detector multiplicity does not improve the sensitivity for 
this analysis \citep{atti03}.}event data have been binned into separated spectra at 0.5 to 50 keV per bin. 

\begin{figure}   
\centering 
\includegraphics[width=6cm,angle=90]{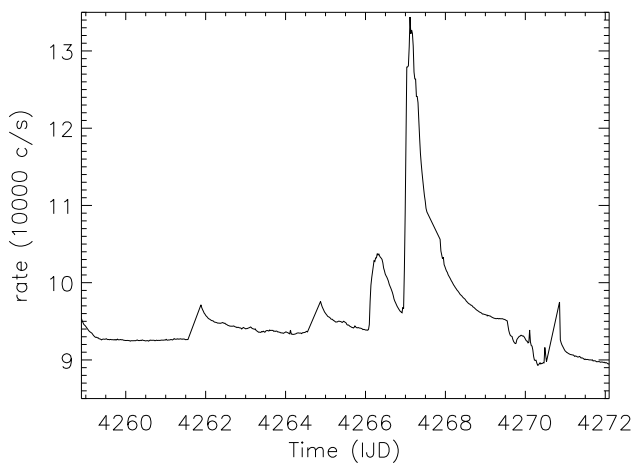}  
\caption{ Time variation of the event rate in the ACS of SPI during the  pre--maximum observations of SN2011fe. }         
\label{spi1}   
\end{figure}

\begin{figure}   
\centering
 \includegraphics[width=6cm,angle=90]{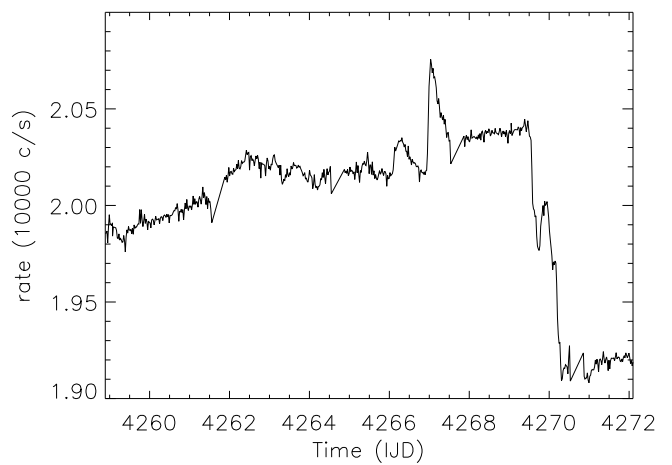}   
\caption{Time variation of the event rate in the GeD of SPI during the pre--maximum observations of SN2011fe. }        
\label{spi2}   
\end{figure}

The time--averaged energy spectrum of the supernova has been extracted by a model fitting method. The flux of the source and the instrumental background are both fitted to the data (counts per detector per pointing at each energy bin), assuming a point source at the SN2011fe position. The instrumental background is fitted per pointing assuming that the relative background rate in each 
detector of the Ge camera is fixed, and is obtained by summing the counts per detector of all 
the pointings of the observation. We have verified that the detector pattern 
did not change with time. This background modelling method is adapted to the 
analysis of data that show strong instrumental background variations 
with time scales $\la 3$ days\footnote{The analyses performed 
with background models that use background tracers show strong 
systematics due to these large instrumental background variations.}. 
Figures \ref{spi1} and \ref{spi2} display the 
count rate in the anticoincidence system (ACS) and in the germanium detector 
(GeD) of SPI. The ACS count rate could be used to trace the 
instrumental background fluctuations in the germanium detectors \citep{jean03}.  The passage through the radiation belts and the impact of the 
solar activity are clearly seen. Solar activity was particularly influential
during this period of observations because of the occurrence of two solar 
flares on $\sim$ 4266 and 4267 IJD followed by a Forbusch decrease around 
4268.5 IJD. In order to check that these background variations do not produce 
artifacts, we also performed the model fitting analysis using a filtered data 
set without periods with strong rate variations. The results obtained with 
filtered and non--filtered data are statistically equivalent. Consequently, 
we decided to use non--filtered data in the next steps of the analysis. 

\begin{figure}  
 \centering
 \includegraphics[width=6cm,angle=-90]{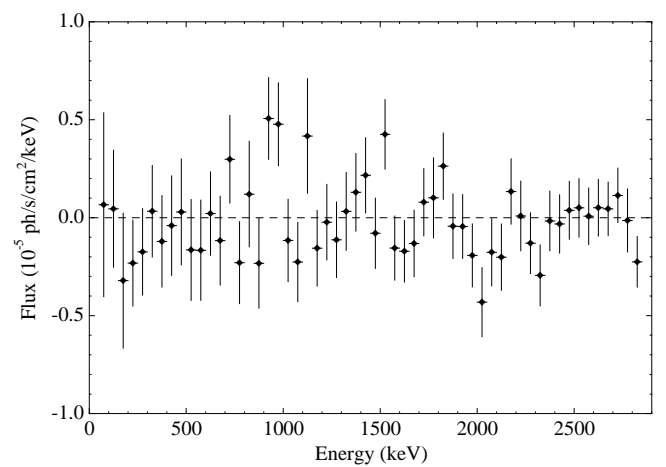}      
\caption{Spectrum of the whole early observation of SN2011fe extracted 
from SPI data with 50 keV bins. }         
\label{spi3}   
\end{figure}

Figure \ref{spi3} displays the spectrum of the whole observation. It has been obtained by combining the spectra extracted by model fitting from single-detector and multiple-detector events. In this case, the model fitting has been performed with data rebinned in 50 keV bins.The spectrum does not show any significant feature. A $\chi^{2}$ test shows that it is consistent with a Poissonian background ($\chi^{2}$ = 12.4 with a dof = 10). A similar conclusion is obtained when the spectra are extracted by model fitting with data rebinned in 5 keV or 2 keV bins.

No significant excess was found in the spectrum, even at the energies of the strongest gamma-ray lines that are expected from the decay of \element[ ][56]{Ni} (blueshifted 158 and 812 keV lines). Table \ref{table:1} presents the upper-limit fluxes derived from the analysis, at the energies of interest for several band widths \citep{iser11}. 

\begin{table}
\caption{Upper-limit of the flux in selected spectral regions for SPI (2$\sigma$), JEM--X (2$\sigma$), and IBIS/ISGRI (3$\sigma$) for the entire pre--maximum observation period.}             
\label{table:1}      
\centering                  
\begin{tabular}{c c c}        
\hline\hline                 
Energy band  & Upper-limit flux                       & Instrument \\ 
            (keV) & (photons s$^{-1}$ cm$^{-2}$) &                   \\  
 \hline                        
3 - 10           &  $5.0 \times 10^{-4}$  & JEM-X      \\
10 - 25         &  $4.0 \times 10^{-4}$  & JEM-X      \\
3 - 25           &  $1.0 \times 10^{-3}$  & JEM-X      \\ 
60 - 172       &  $1.5 \times 10^{-4}$  & IBIS/ISGRI \\
90 - 172       &  $1.1 \times 10^{-4}$  & IBIS/ISGRI \\
150 - 172     &  $7.1 \times 10^{-5}$  & IBIS/ISGRI \\    
160 - 166     &  $7.5 \times 10^{-5}$  & SPI            \\     
140 - 175     &  $2.3 \times 10^{-4}$  & SPI            \\     
814 - 846     &  $2.3 \times 10^{-4}$  & SPI            \\     
800 - 900     &  $3.5 \times 10^{-4}$  & SPI             \\     
\hline                                   
\end{tabular}
\end{table}

\subsection{IBIS/ISGRI data}

IBIS uses two detection layers to cover the same energy range as SPI. 
The low energy camera, ISGRI, uses 16,384 thin CdTe detectors operated 
at ambient temperature. The imaging is much better ($12\arcmin$ FWHM) 
than for SPI but the spectral resolution is more limited and the efficiency begins 
to drop above 100 keV. As a result, the IBIS/ISGRI sensitivity to the 
812~keV line from \element[][56]{Ni} is far worse than the SPI one at 
the same energy even in the case of a broad line. Conversely, 
IBIS/ISGRI is much better at work than SPI to reveal a low energy continuum 
or broad lines such as the 158 keV \element[][56]{Ni} or 122~keV \element[][57]{Co} 
ones. To summarize the picture, one could say that SPI is better at hand 
to detect narrow lines in a spectrum while IBIS/ISGRI is better at detecting 
point sources in broad energy-range sky images. Two data processing methods 
have been followed in parallel. One uses the standard OSA--9 version and the 
other takes advantage of the developments for the forthcoming OSA--10. The latter approach includes two new corrections: for the spectral drift along the mission and for the flat field. In either case no significant (greater than $3\,\sigma$) signal was found at the position of SN2011fe. Table \ref{table:1} gives the upper limits obtained for the 60-170 keV, 90-172 keV and the 150-172 keV bands. Similarly, maps using the last 6 days of the observing period for the same energy ranges  do not show  any significant emission at the position of SN2011fe.

\subsection{JEM--X data}
Both JEM-X units were simultaneously operating at the time of these 
\textit{INTEGRAL} observations and they can be used to constrainthe continuum emission using a source search in broad band images as in the IBIS/ISGRI case. Due to the smaller field of view of 
the JEM-X monitors compared to those of SPI and IBIS \citep{lund03}, 
we selected and analyzed only those pointings where SN2011fe was
within 5$^\circ$ of the pointing direction. The data obtained during 
the orbits 1084 -- 1088 (August 29th to September 12th, 2011) were 
analysed with OSA-9 following the standard procedure. Images from single 
pointings were combined into one mosaicked image for each X-ray monitor. 
The two composite images from JEM-X1 and JEM-X2 were then merged to obtain 
the final image, providing an on-source effective exposure time of 450 ks. 
The imaging analysis was performed in the 3--25~keV band, and in the 3--10~keV 
and 10--25~keV sub-bands. SN2011fe is not detected in any of the JEM-X images. 
Assuming a Crab-like spectrum, we estimate $2 \, \sigma$ upper limits on the 
3--10~keV and 10--25~keV fluxes of 
$5 \times 10^{-4} \, \rm ph \, cm^{-2} \, s^{-1}$ and $4 \times 10^{-4} \, \rm ph \, cm^{-2} \, s^{-1}$, 
respectively (see Table \ref{table:1}). 
The source is also not detected when analyzing separately the JEM-X data 
from orbits 1086 -- 1088 (Sept 4th to 12th, 2011), corresponding to the 
observations closer to the epoch of the expected $^{56} \rm Ni$ line maximum.

\section{Discussion}
The $\gamma$--ray spectrum of SNIa mainly depends on the total amount and distribution of
\element[][56]{Ni} synthesized during the explosion, as well as on the chemical structure and
velocity distribution of the debris \citep{gome98}. For instance, when \element[][56]{Ni} is
present in the outer layers of some of the sub--Chandrasekhar models, the corresponding lines
should appear very early in the spectrum. However, for a given flux, there are at least two more
factors that determine if the $\gamma$--signal is detectable: the change of 
the signal with time and the width of the lines.  
  
\begin{table}
\caption{Kinetic energy (K) and mass of \element[ ][56]{Ni} produced by different models of explosion.}             
\label{table:2}      
\centering                  
\begin{tabular}{c c c}        
\hline\hline                 
Model  & $K$ (foe) &  $M_\mathrm{Ni}$  (M$_\odot$)                       \\  
 \hline                        
DETO     &    1.44   & 1.16      \\
DD202c &    1.30   & 0.78      \\
DDTc     &    1.16   & 0.74      \\
SC3F      &    1.17   & 0.69      \\
W7         &    1.24   & 0.59      \\
SCOP3D &    1.17   & 0.56      \\
DDTe     &    1.09   & 0.51      \\
SC1F      &    1.04   & 0.43      \\
HED6     &    0.72   & 0.26       \\ 
\hline                                   
\end{tabular}
\end{table}

In order to check the influence of the mass and distribution of \element[ ][56]{Ni}, several models
obtained under different hypothesis about the burning regime or the explosion mechanism have been considered. 
Most are one-dimensional spherically symmetric models, which should be sufficient for SN2011fe in view of the small 
amount of global asymmetry suggested by spectropolarimetric measurements \citep{smi11}. 
We have not included models of DD explosions for two reasons. First, \citet{nuge11a} 
reproduced satisfactorily the rising part of the light curve of SN2011fe with a simple 
analytic model involving a Chandrasekhar mass WD, and found only small amounts of unburned carbon in the early spectra of this supernova. Second, in spite of recent advances on simulations of almost normal SNIa from mergers of massive WDs \citep{pak12}, theoretical models of DD explosions are not as mature at present as those involving either Chandrasekhar or sub-Chandrasekhar mass WDs accreting from a non-degenerate companion. Hence, the number of free parameters involved in DD explosions is too large to allow for efficient constraints derived from upper limits on the gamma-ray emission of SN2011fe. 

Table \ref{table:2} displays the main characteristics of the models used in 
the present study. The DETO model \citep{bade03} corresponds to a pure detonation of a WD near the
Chandrasekhar's limit. It is also representative of the most massive models computed by
\cite{fink10}. The W7 is the classical model of \cite{nomo84}. The DD202c \citep{hoef98} and the
DDTc,e \citep{bade05} models are delayed detonation models that produce different amounts of
\element[ ][56]{Ni} and have different expansion energies. The HED6 model corresponds to the
explosion of a 0.6 M$_\odot$ C/O WD that has accreted 0.17 M$_\odot$ of 
helium of \cite{hoef96}, and SC1F and SC3F (E.Bravo, unpublished) are 
also sub-Chandrasekhar models equivalent to models 1 and 3 of \cite{fink10}. 
Finally, the SCOP3D model is a three dimensional sub--Chandrasekhar model that corresponds to model A of \cite{garc99}.

\begin{figure}   
\centering   
\includegraphics[width=9cm]{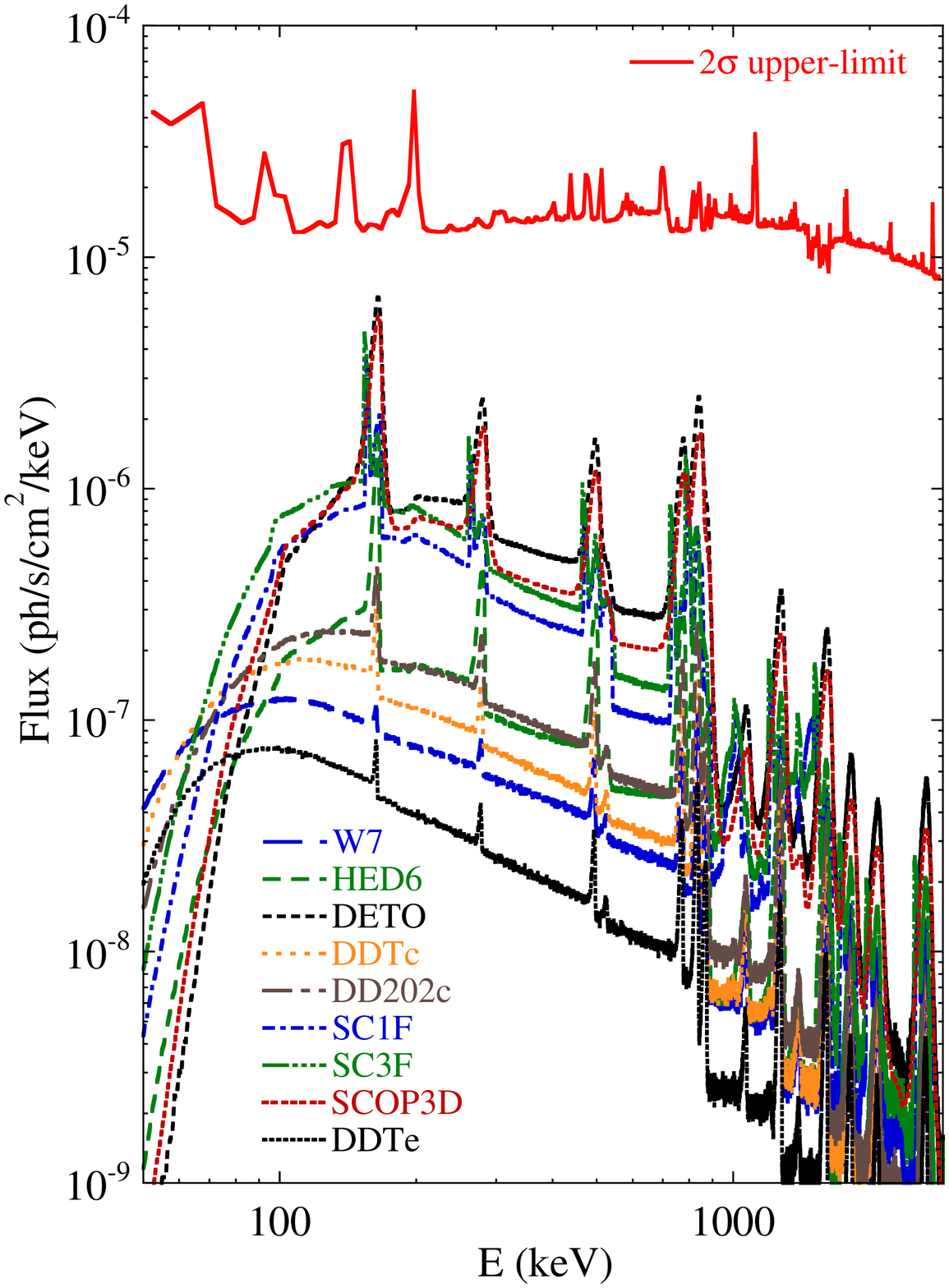} 
 \caption{Comparison of the mean flux obtained from the theoretical models during the early observation period with the  2$\sigma$ upper-limit flux in 5 keV bins. Theoretical spectra were obtained with the code described in \cite{gome98}.}	 
\label{spi4}   
\end{figure}   

\begin{table*}   
\caption{Measured and predicted flux in spectral regions where \element[ ][56]{Ni} lines are expected to be as a function of models. The fifth column is the probability to reject the model (see text). In the case of DDTc, W7 and DDTe models, the fluxes were obtained for an optimal 
observation periods of 7, 5 and 5 days, respectively (see Table\ref{table:4}).}               
\label{table:3}   \centering                          
 \begin{tabular}{c c c c c}       
 \hline\hline         
 Model     &    Energy band &         Measured flux &   Predicted flux   & Probability \\  
                &            (keV)      & (10$^{-4}$ photons s$^{-1}$ cm$^{-2}$) & (10$^{-4}$
photons    s$^{-1}$ cm$^{-2}$) & \%\ \\       
\hline          
    DETO & 150 - 175  &  -0.20 $\pm$ 0.78  &  0.81  &  92.4   \\       
              & 265 - 295  &  -0.55 $\pm$ 0.85  &  0.48  &            \\  
              & 482 - 508  &    0.13 $\pm$ 0.88  &  0.36  &            \\
              & 730 - 880  &  -1.01 $\pm$ 2.02  &  2.01  &            \\        
\hline           
   SCOP3D & 145 - 170  &  -0.50 $\pm$ 0.76  &  0.70  &  84.7   \\ 
              & 267 - 297  &  -0.24 $\pm$ 0.86  &  0.36  &            \\ 
              & 482 - 508  &    0.13 $\pm$ 0.88  &  0.27  &            \\
              & 720 - 880  &  -0.75 $\pm$ 2.08  &  1.54  &            \\        
\hline              
     SC3F & 150 - 170  &  -0.03 $\pm$ 0.67  &  0.37  &  62.4   \\ 
              & 253 - 283  &  -0.99 $\pm$ 0.85  &  0.25  &            \\
              & 452 - 508  &    0.38 $\pm$ 1.42  &  0.29  &            \\
              & 720 - 880  &  -0.75 $\pm$ 2.08  &  0.76  &             \\       
 \hline                    
     SC1F & 150 - 170  &  -0.03 $\pm$ 0.67  &  0.32   &  51.8   \\ 
              & 254 - 288  &  -0.81 $\pm$ 0.90  &  0.22   &            \\ 
              & 458 - 509  &    0.01 $\pm$ 1.37  &  0.23   &            \\
              & 725 - 900  &  -1.02 $\pm$ 2.18  &  0.63   &            \\        
\hline                    
     HED6 & 159 - 169  &  -0.13 $\pm$ 0.49  &  0.08   &  21.3   \\ 
	   & 270 - 284  &  -0.98 $\pm$ 0.59  &  0.05   &            \\ 
	   & 475 - 505  &  -0.27 $\pm$ 1.04  &  0.06   &            \\
	   & 746 - 872  &  -0.13 $\pm$ 1.87  &  0.27   &            \\ 
\hline  
     DDTc & 158 - 165  &  -0.10 $\pm$ 0.57  &  0.03   &  6.2   \\ 
	  & 270 - 282  &  -0.58 $\pm$ 0.76  &  0.02   &            \\ 
	  & 480 - 498  &  -1.22 $\pm$ 1.00  &  0.02   &            \\
	  & 740 - 880  &  1.28 $\pm$ 2.08  &  0.08   &            \\
\hline   
     W7 & 158 - 165  &  -0.34 $\pm$ 0.69  &  0.02   &  3.4   \\ 
	  & 270 - 282  & -0.88 $\pm$ 0.92  &  0.02   &            \\ 
	  & 478 - 498  & -1.82 $\pm$ 1.37  &  0.02   &            \\
	  & 740 - 880  &  -0.31 $\pm$ 1.38  &  0.04   &            \\
\hline                                   
     DDTe & 158 - 165  &  -0.34 $\pm$ 0.69  &  0.01   &  2.0   \\ 
	 & 270 - 280  &  -0.49 $\pm$ 0.84  &  0.008   &            \\ 
	 & 480 - 498  &    -1.00 $\pm$ 1.26  &  0.009   &            \\
	 & 740 - 880  &  -0.152 $\pm$ 3.25  &  0.05   &            \\        
\hline                                   
\end{tabular}   
\end{table*}

Figure \ref{spi4} displays the SPI 2$\sigma$ upper-limit spectrum obtained with 5 keV bins as well as the gamma ray spectra predicted by several models for this early observation, assuming that the distance of SN2011fe is 6.4 Mpc. All the models are well below the upper--limit. The predicted intensity of the  \element[ ][56]{Ni} 158 keV, 270 keV, 480 keV, 750 keV, and 812 keV lines is maximum for the detonation model (DETO) and the Sub-Chandrasekhar's models SCOP3D, SC3F, and SC1F, as expected. Since the spectral shape (width and centroid) and intensity change from one model to another, the energy band used to extract the fluxes of every line or complex of lines was chosen to provide the optimum significance for each model. Results are displayed in Table \ref{table:3}.

Despite the non-detection and the bounds imposed by the optical light curve, the compatibility of the different models with the zero flux hypothesis has been tested. Since the measured fluxes are compatible with zero, the $\chi^{2}$ value for each model was computed as the quadratic sum of the expected significance of the predicted fluxes\footnote{The expected significance is the predicted flux divided by the uncertainty of the flux measurement in the corresponding energy band.}. Then, the probability of rejection of each particular model was derived by computing the probability that the $\chi^{2}$ is smaller than the measured value within $\chi^{2}$ statistics. The rejection probability of the selected models is given in the fifth column of Table \ref{table:3}. None of them can be firmly rejected by SPI data. Although the
probability to reject the DETO model is 92.4\%\ this value corresponds to a significance of
$\sim$1.4$\sigma$. To reject it at a 99\% of confidence level, the intensity of the lines should
be larger by a factor 1.45 times or, equivalently, the distance to SN2011fe should be less than 5.3
Mpc. Since the DETO model is the one that produces the brightest lines, SPI can only provide
interesting results during this epoch if the distance to the supernova is smaller than this value.

The influence of the temporal behavior of the line intensity and line width on the detectability of
the $\gamma$ emission can be easily seen by estimating the significance of the observation. In the limit of weak signals this significance is given by \citep{jean96}:

\begin{equation}
n_{\sigma} = \frac{{A_{eff} \int\limits_{t_i  - \Delta t}^{t_i } {\varphi \left( t \right)dt} }}{{\sqrt {bV\Delta E \Delta t} }} 
\label{nsigma}
\end{equation}
\noindent where $\Delta t$ is the observation time, $A_{eff}$ th​e effective area at the
corresponding energies, $\varphi$ is the flux  (cm$^{-2}$s$^{-1}$) 
in the energy band $\Delta E$,  V is the volume of the detector and b is the 
specific noise rate (cm$^{-3}$s$^{-1}$keV$^{-1}$), where it has been assumed 
that it is weakly dependent of the energy and time in the interval of interest.

 If the flux grows like $ \varphi \left( t \right) = \varphi _0 e^{\alpha t} $, the significance
 reached by integrating the time interval $\left( {t_i  - \Delta t,t_i } \right)$ is
\begin{equation} 
 n = \frac{{A_{eff} \varphi \left( {t_i } \right)}}{{\sqrt {\alpha bV\Delta E} }}\frac{{1 - e^{ -
\alpha \Delta t} }}{{\sqrt {\alpha \Delta t} }} 
\end{equation}
 \noindent For $\alpha \Delta t <  < 1$, it behaves as $n \propto \sqrt{\Delta t}$ and has a
maximum at $ \alpha \Delta t = 1 .26$. Furthermore, the dependence on  $\Delta E$, clearly shows the convenience of taking a value that maximizes the signal/noise ratio. Unfortunately, since the value of $\alpha$ is not known \emph{a priori} the optimal observing time is not known in advance. Since the time dependence of the model fluxes does not follow strictly an exponential growth, we have analysed the optimum observation periods for each model. Table~\ref{table:4} displays the results. In general, models and energy ranges showing weak variation of the flux are best detected when the observation period is large whereas models with a strong variation of their flux during the first two weeks are best detected using data at those times.

\begin{table}   
\caption{Optimal observation period for several models and energy ranges with SPI.}               
\label{table:4}   \centering                          
 \begin{tabular}{c c c }       
 \hline\hline         
 Model     &    Energy band &      Optimal period \\  
                &            (keV)      & (days) \\       
\hline          
    DETO & 150 - 175  &  13.5   \\       
         & 730 - 880  &  13.5  \\        
\hline           
    SCOP3D & 145 - 170  &  13.5   \\ 
           & 750 - 880  &  13.5   \\        
\hline              
    SC1F & 150 - 170  &  13.4  \\ 
         & 725 - 900  &  13.4   \\        
\hline              
    HED6 & 155 - 175  &  11.7  \\ 
         & 730 - 880  &  12.4   \\        
\hline
    DDTc & 70 - 165 & 8.0 \\
         & 740 - 880 & 6.2 \\
\hline 
    W7 & 70 - 900 & 5.7 \\
       & 820 - 840 & 4.7 \\
\hline 
    DDTe & 158 - 165 & 4.9 \\
         & 740 - 880 & 4.9 \\
\hline                                   
\end{tabular}   
\end{table}

It is well known \citep{gome98} that the width of the line has a strong influence on their
detectability. In the case of the present observations it is possible to estimate the significance
of a narrow line, $\Delta E \la 2$ keV, by comparing the measured $1\sigma$ flux uncertainty (see Table \ref{table:3}), $\varphi_1$, with the flux $\varphi$ and the width 
$\Delta E$ predicted by the model

\begin{equation}
n_\sigma \approx \frac{\varphi}{\varphi_1}\sqrt{\frac{\Delta E}{2\,\,{\rm keV}}}
\end{equation}

\noindent In the case of the 158 keV line (Table \ref{table:3}), the flux predicted by the DETO
model is $\varphi = 8.1 \times 10^{-5}$ ph s$^{-1}$cm$^{-2}$, $\varphi_1 = 7.8 \times 10^{-5}$ ph s$^{-1}$cm$^{-2}$, and $\Delta E =25$ keV, where $\Delta E$ has been chosen to maximize the signal/noise ratio. With these data, $n \sim 3.8$ and in this case, with this hypothesis, the 158 keV line would have been detected by SPI if it had been narrow. On the contrary, in the model SC1F, $\varphi = 3.2 \times 10^{-5}$ ph s$^{-1}$cm$^{-2}$, $\varphi_1 = 6.7 \times 10^{-5}$ ph s$^{-1}$cm$^{-2}$, $\Delta E =20$ keV and the result is $n\sim 1.5$. Here, the line would only be marginally detectable even in the narrow case.

   \begin{figure}
   \centering
   \includegraphics[width=9cm]{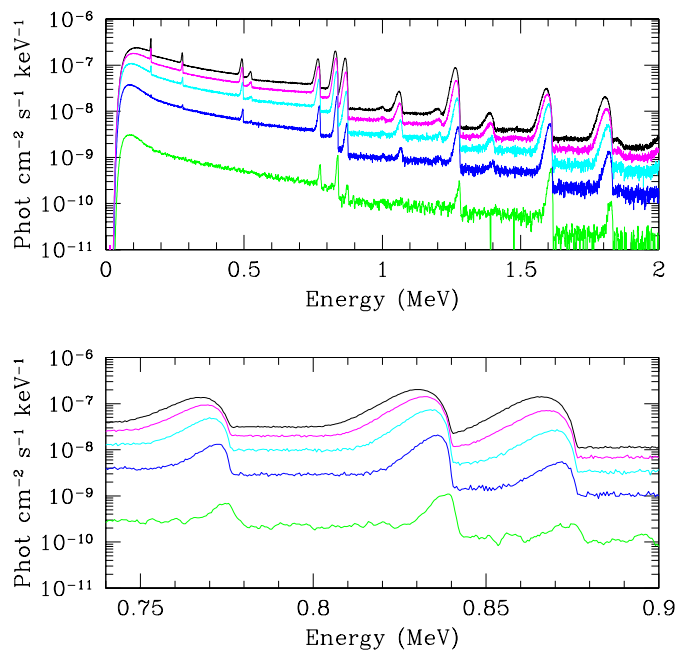}
      \caption{Early gamma ray spectra predicted by the delayed detonation model that best fits the OMC (see Fig. \ref{omc2}) at days 8, 11, 14, 17, 20 (green, blue, cyan, magenta, black) at a distance of 6.4 Mpc (upperer figure). The lower pannel displays the evolution of the profile of the \element[ ][56]{Ni} 812 kev--\element[ ][56]{Co} 847 keV features.
              }
         \label{specDDTe}
   \end{figure}

The luminosity at the maximum of the optical light curve is proportional to the total mass of
\element[ ][56]{Ni}, while the shape of the optical peak  is determined by the amount of \element[
][56]{Ni} and the kinetic energy of the remnant \citep{arne97}. As it has already been mentioned,
the favoured model by the optical light curve is the DDTe one, and it seems worthwhile to interpret the SPI observations in terms of this model.

Figure \ref{specDDTe} (upper panel)  displays the early gamma ray spectra predicted by DDTe for
several instants after the explosion. The way the different lines grow and the width of them is clearly seen. The main problem with the 812 keV line is that, due to the Doppler effect 
and energy degradation of photons by Compton scattering, it blends with the 847 keV line of 
\element[ ][56]{Co} (see Fig. \ref{specDDTe}, lower panel) and makes more
difficult the interpretation of the signal. During this observation, and for this model, the
energy interval that maximizes the signal/noise ratio is $\sim 25$ keV centered at 826 keV, 
and the corresponding flux at day 18 is $\varphi = 3.4 \times 10^{-6}$ ph s$^{-1}$cm$^{-2}$. 
On the contrary, the 158 keV line has a more regular behavior and it is better suited for diagnostic purposes. At day 20 after the explosion, the flux in this line is $\varphi = 4.7 \times
10^{-6}$ ph s$^{-1}$cm$^{-2}$ for $\Delta E =20$ keV. In both cases, the signal is too weak 
to be detected with SPI and the source should be at a distance of $\la 2$ Mpc to be detectable.



The 200--540 keV band contains almost all the photons produced by the annihilation of positrons produced during the \element[ ][56]{Ni} chain decay. Since the emission of 511 keV only represents the 25\% of the total number of the annihilation photons, this band offers, in principle, better possibilities of detection than the 511 keV line itself. In the DDTe model, the maximum of the band occurs at day 50 and amounts $1.05 \times 10^{-4}$ ph s$^{-1}$cm$^{-2}$, while the maximum of the 511 keV line occurs at day 74 with  a flux of $1.2 \times 10^{-5}$ ph s$^{-1}$cm$^{-2}$, within a band of 30 keV. The expected emission at day 18 after the explosion, the last day of this observation window, are $4.9 \times 10^{-5}$ ph s$^{-1}$cm$^{-2}$ and $2.4 \times 10^{-6}$ ph s$^{-1}$cm$^{-2}$ respectively. The $3\,\sigma$ sensitivity in this band is estimated to be $8.7 \times 10^{-4}$ ph s$^{-1}$cm$^{-2}$ for an integration time of $10^6$ s, well above the value of the expected emission for this model, for which reason this band is not detectable by SPI/\emph{INTEGRAL} except if the explosion occurs at a distance $d\la 1-2$ Mpc (assuming an explosion similar to that predicted by the DDTe model).

\begin{figure}
   \centering
 \includegraphics[width=9cm]{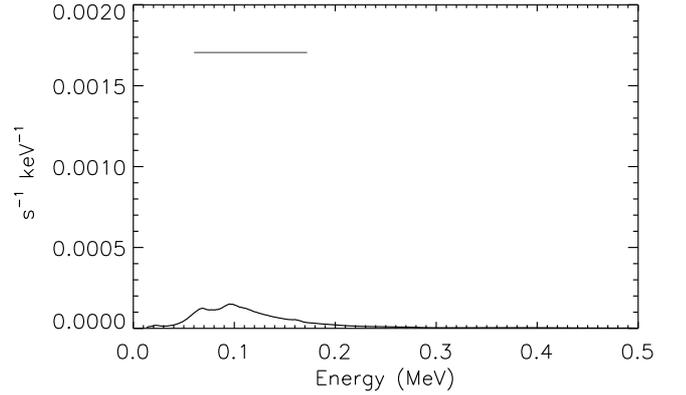}
      \caption{ISGRI expected rate--spectrum from a DDTe model averaged over the last 6 days of the first observation period. The $3\,\sigma$ ISGRI upper limit for the 60--172 keV band is displayed as an horizontal bar.
              }
         \label{isgri2} 
   \end{figure}

As in the case of SPI, it is possible to discuss the results from ISGRI in more depth using detailed predictions by different theoretical models. In this case we use the model to produce a daily spectrum corresponding to the period of observation and combine these spectra with the ISGRI sensitive area (ARF) and spectral response (RMF) to predict the observable ISGRI spectrum as a function of time. Using the latest ISGRI background estimate (F. Mattana, private communication), the expected relative value of signal to noise ratio can be computed as a function of energy and time. This was done for the DDTe, W7, DETO and SC3F models. The first two of these models predict fluxes that are out of reach of the ISGRI sensitivity by an order of magnitude even if only the last days of the observing period are used, where the expected signal is higher. Figure \ref{isgri2} illustrates an attempt in the case of the DDTe model. The 60 - 172 keV upper limit on the ISGRI rate during the last six days of the first observation period is $1.7 \times 10^{-3}$ s$^{-1}$keV$^{-1}$, which represents $2.2 \times 10^{-4}$ cm$^{-2}$s$^{-1}$.


   \begin{figure}
   \centering
 \includegraphics[width=9cm]{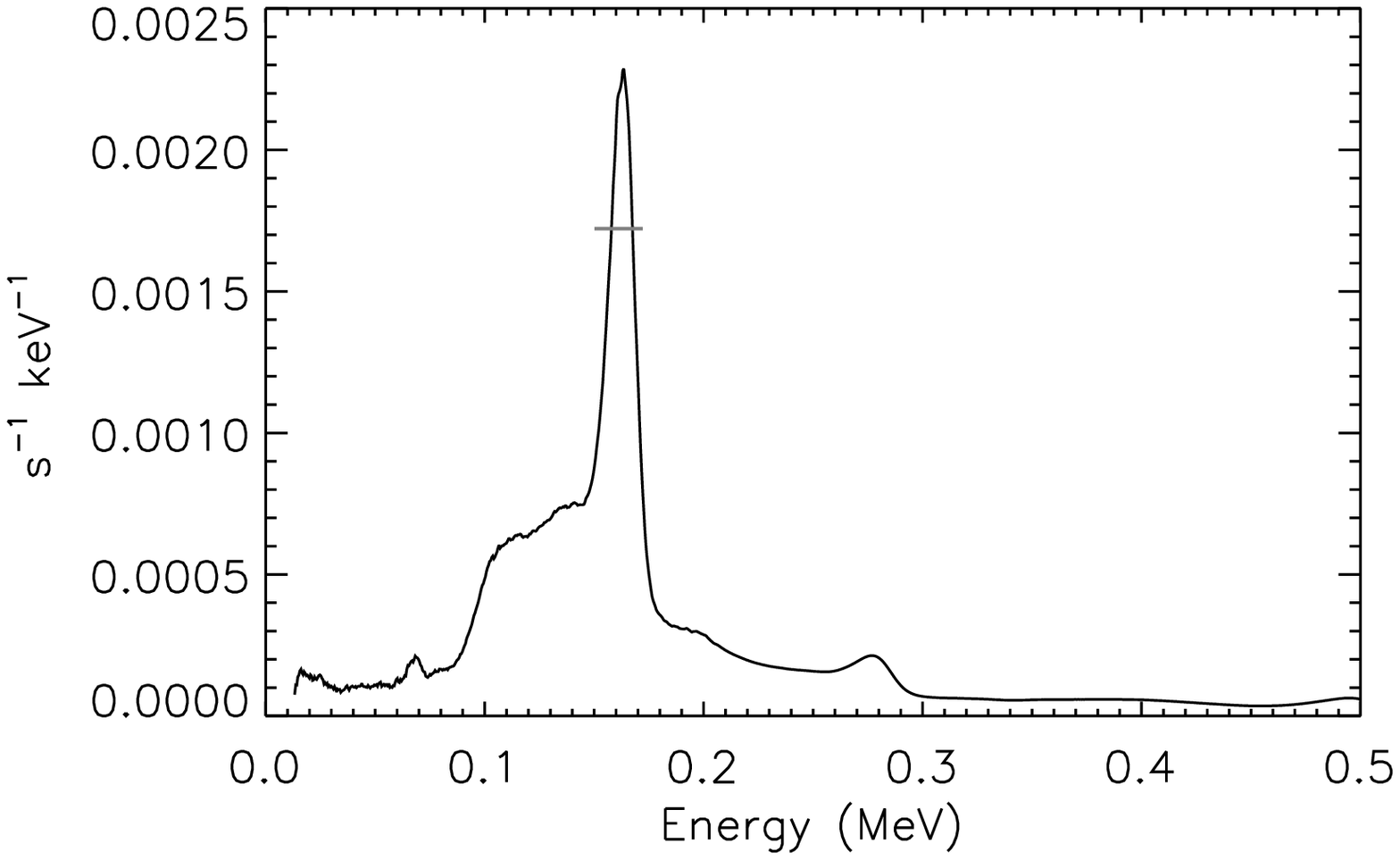}
      \caption{ISGRI expected rate--spectrum from a DETO model averaged over the entire observation period. The $3\,\sigma$ ISGRI upper limit for the 150--172 keV band is displayed as an horizontal bar.
              }
         \label{isgri3}
   \end{figure}
 
   \begin{figure}
   \centering
 \includegraphics[width=9cm]{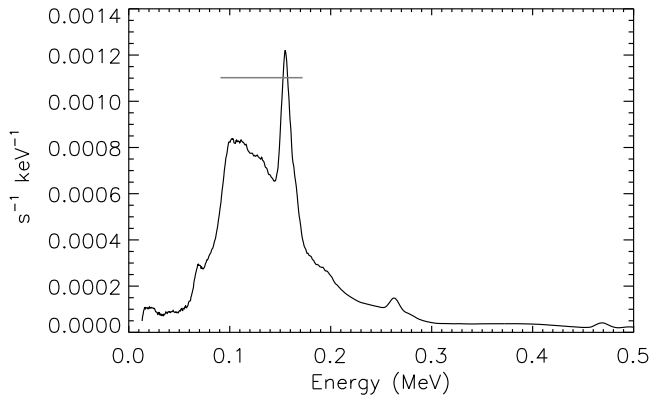}
      \caption{ISGRI expected rate--spectrum from a SC3F model averaged over the entire observation period. The $3\,\sigma$ ISGRI upper limit for the 90--172 keV band is displayed as an horizontal bar.
              }
         \label{isgri4}
   \end{figure}

Since the DETO and SC3F models produce important amounts of \element[ ][56]{Ni} near the surface, they display an important emission feature in the  $\sim 150 - 170$ keV band. Figure \ref{isgri3} displays the spectrum that could be expected from the DETO model. 
This spectrum is an average over all the time that ISGRI has been pointed to the source and the optimal band corresponds 
to 150 - 172 keV.  In this case the $3\,\sigma$ limit corresponds 
to $7.1 \times 10^{-5}$ cm$^{-2}$s$^{-1}$. If a source like DETO 
would have been present, the significance of the signal would have 
been $2.8 \,\sigma$, but there was nothing detected by ISGRI in the field. 
Model SC3F exhibits a similar behavior (figure \ref{isgri4} ) and, 
for the same specifications, the significance in the band is 
$1.4 \,\sigma$. However, in this case, the continous emission at low energies 
is important and the optimal observation band is 90 - 172 keV, the $3\,\sigma$ 
limit is $1.1 \times 10^{-4}$ cm$^{-2}$s$^{-1}$ and the significance of 
the signal would have been $2.1 \,\sigma$. Notice that if these models, 
that synthesize 1.16 and 0.69 M$_\odot$ of \element[ ][56]{Ni} respectively, are scaled to the $\sim 0.5$ M$_\odot$ demanded by the observations of SN2011fe in the optical, their expected gamma-ray flux would be marginally above the ISGRI sensitivity limit. 

\section{Conclusions}
SN2011fe has been observed with the four instruments SPI, ISGRI/IBIS, JEM-X and OMC 
operating on board of \emph{INTEGRAL} just before the maximum of the optical light 
curve for a period of time of 1000 ks and the data were compared with the 
predictions of several theoretical models. SPI data in the bands containing 
the 158~keV and the 812~keV emission of \element[ ][56]{Ni} would have allowed to reject at 98\% confidence level ($2\sigma$) models that produce large amounts of  \element[ ][56]{Ni} near the surface, if the supernova were closer. For instance, models with a mass fraction of \element[ ][56]{Ni} in the outer 0.15 M$_\odot$ of 0.05 and 0.5 would have been rejected if the supernova distance was closer than 1.4 and 3.7 Mpc  respectively, assuming non-detection. 
Furthermore, ISGRI/IBIS has proved to be very efficient to explore the low energy 
region ($ \sim100$ keV) and has confirmed that there were not significant amounts 
of radioactive elements in the outer layers. This picture is consistent with the 
light curve obtained with the OMC. The observations in the optical suggest that the 
total amount of \element[ ][56]{Ni} produced in the event is 
$\sim 0.5$ M$_\odot$ originated by a mild delayed detonation explosion.

\begin{acknowledgements}
This  work has been  supported by the  MINECO-FEDER grants AYA08-1839/ESP, AYA2011-24704/ESP, AYA2011-24780/ESP, AYA2009-14648-C02-01, CONSOLIDER CSC2007-00050,  by the ESF EUROCORES Program EuroGENESIS  (MINECO grants  EUI2009-04170), by the grant 2009SGR315 of the Generalitat de Catalunya. In parts, this work has also been supported by the NSF grants AST-0708855 and AST-1008962 to PAH.

The INTEGRAL SPI project has been completed under the responsibility and leadership of CNES.
We also acknowledge the INTEGRAL Project Scientist Chris Winkler (ESA, ESTEC) and the INTEGRAL personnel for their support to these observations.
\end{acknowledgements}

\bibliographystyle{aa}
\bibliography{sn2011fe_1}
\end{document}